# Impact of the Digital Transformation: An Online Real-Time Delphi Study

Florian Cech, C!S, TU Wien
Hilda Tellioğlu, C!S, TU Wien

**Abstract.** The effects of the ongoing Digital Transformation of society are far-reaching and subject to a multitude of impact factors. Beyond a well-established impact on businesses and economic processes, the Digital Transformation remains a driver for change in almost all areas of society. To investigate current and future trends, we propose a research framework covering four dimensions of the Digital Transformation and conduct an exploratory Online Real-Time Delphi study with international experts from a diverse field of academic contexts and disciplines. Our study indicates the societal areas most likely to be affected by the Digital Transformation, the relevant issues and global challenges connected to and influenced by the Digital Transformation, and the impact of key digital technologies now and in the next five years. Given these results, we evaluate the research framework and our approach to investigate the most relevant aspects of the Digital Transformation.

# 1 Introduction

There are only a few technological shifts as impactful and consequential for society on a global scale as the advent of digital technologies. Within the relatively short time span of about two and a half decades, technological advancements like the Internet, the widespread use of smart phones or social media, to name only a few, have thoroughly influenced and – arguably – changed society: hence the term *Digital Transformation*. Far from being complete, the accelerating pace of technological advancement continues to disrupt, change and transform established norms and practices throughout many societal areas. The exact nature of these transformations is difficult to pinpoint and even harder to predict, since many of the involved technologies carry both the potential to greatly benefit society and the risk of significantly harmful effects [16, pp. 4-6]. These difficulties notwithstanding, researchers from a multitude of fields and disciplines have studied a variety of specific aspects of the Digital Transformation, but only rarely have these attempts taken on a macroscopic, interdisciplinary view.

This study attempts to take this macroscopic stance, first and foremost by framing the problem in a structured way. Following this, an Online Real-Time Delphi Study was conducted to investigate the relationship between the four dimensions of the Digital Transformation as proposed in the framework. Finally, we sum up the results and present a set of recommendations for future research based on the qualitative and quantitative evaluation of the study.

# 2 Framing the Context

The process of the Digital Transformation is, depending on context and impact, sometimes referred to as "Digitalization" or "Digital Disruption". Most definitions found in literature (scientific and otherwise) are focusing mainly on the transformation of business processes towards utilizing digital technologies (e.g. [26, 30, 34, 35]). For instance, Westerman et al. describe it simply as *"[...] the use of technology to radically improve performance or reach of enterprises [...]"* [37], while Schallmo and Williams provide a more comprehensive definition:

> *"[We] define digital transformation as a sustainable, company-level transformation via revised or newly created business operations and business models achieved through value-added digitization initiatives, ultimately resulting in improved profitability."* [34, p.4]

While businesses and economies are certainly one of the areas impacted, the effects of the permeation of society with digital technologies are much more far-reaching than just the disruption of existing business practices. To broaden the scope of this study, we utilize more general definitions that apply to other contexts or societal areas as well. Katz, for instance, describes the Digital Transformation as

> *"[...] the transformations triggered by the massive adoption of digital technologies that generate, process, share and transfer information."* [22, p.4]

While definitions like the one above take a very generic approach, research taking a macroscopic view on the issue of Digital Transformation is scarce and tends to focus on specific issues, areas or scientific fields affected by Digital Transformation. Some investigate digital government strategies and policy with a focus on gender equality, others focus on the concept of Digital Natives or on specific technological approaches like Big Data or Machine Learning ([19,21,24]). From a sociological perspective, we've learned about the effects of the Digital Transformation on human intimacy and sexuality (e.g., [5,9,20]) and investigated the future of work in the context of Digital Transformation (e.g., [4,6,7,15,38]), to name a few examples.

On an even larger scale, the advent of digital technologies and digital media has given way to entire inter- and trans-disciplinary scientific fields, such as the Digital Humanities, Social Informatics or Digital Culture Studies.

In addition to these heterogeneous definitions and research foci in different fields, the attribution of research to this field is not always easy: A number of publications clearly analyse transformational processes caused or influenced by the rise of digital technologies, but don't specifically refer to any of the synonyms for Digital Transformation (e.g., [10,28,33]).

## 3   Framing the Problem

Having illustrated the heterogeneous nature of research into Digital Transformation, conceptualizing the problem in a research framework capable of covering a macroscopic view on the future-oriented questions of this study presents a major challenge. Expanding on Gudergan and Mugge's [14] call to utilize a holistic approach and bridging the gap between technical and social sciences, the framework we propose outlines four conceptual dimensions of the Digital Transformation: *Societal Areas*, *Issues*, *Technologies*, and *Global Challenges* (Figure 1).

*Societal areas* cover a wide range of aspects of society, from *personal life* to *health care*, *education*, *mobility*, or *economy*. The categories are kept broad intentionally, allowing for some overlap to provide opportunities for discussion within the study itself, as well as leaving certain aspects open to interpretation.

As with any paradigm shifts on a larger societal level, the Digital Transformation gives rise to a number of *issues* that challenge its potential positive effects. The ubiquity of digital devices amplifies issues of *privacy* and *computer security*, and transforms the way we see and present ourselves and our *digital identities*. Changes in the way we produce and consume news require changing how we gauge the veracity of media content, and striving for equality must now include addressing a growing and transforming *Digital Divide* that leaves vulnerable members of society behind. Finally, an economy that is increasingly based on automation, be that in manufacturing, logistics or even the service industry, must provide answers to issues of *reduced employment opportunities* and *job security*.

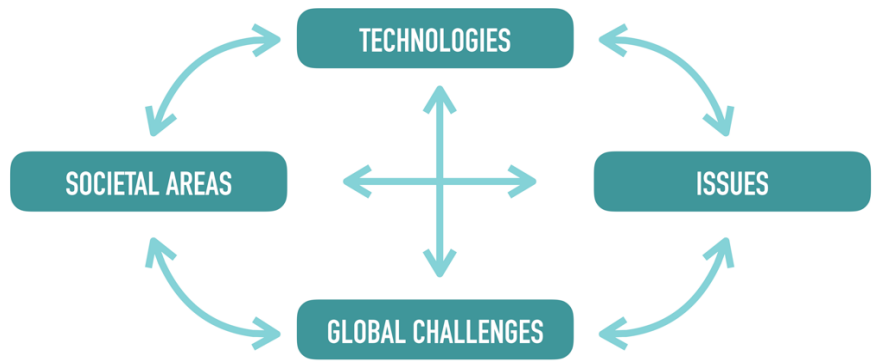

**Figure 1.** Four Dimensions of the Digital Transformation

Looking beyond local issues, humanity faces a number of *global challenges* that will require cooperation and sustained, combined efforts to solve. The Digital Transformation carries the potential to provide solutions to these challenges, but also amplify the related issues: while digital technologies can help battle climate change, advance equality and improve human rights, they can also have a potentially adverse impact on geographic regions affected by poverty, war or economic underdevelopment.

Finally, core aspects are the *(digital) technologies* advancing the Digital Transformation of society. Some technologies, like the Internet itself, have had time to mature, while others are comparably new, making their future impact on society a difficult matter to predict. Not all of these technologies will prove to have a lasting impact or even relevance, but they all carry the potential to broadly affect our daily lives in one way or another.

These four dimensions lead to the following four main research questions:

- Which *societal areas* will be impacted most and least by the Digital Transformation in the next five years?
- Which *issues* related to the Digital Transformation are the most and least important to address within the next five years?
- What impact will the Digital Transformation have on *global challenges* in the next five years?
- Which *technologies* will have the most and least impact in the next five years?

The four dimensions and the relative ranking of items within them provide only the first part of the investigative insights into the Digital Transformation.

As such, the research framework strives not for completeness or comprehensiveness, but as a facilitator for sense making – allowing investigators of the Digital Transformation to come to *"[...] a meaningful and functional representation of some aspects of the world."* [23] – within the Delphi methodological approach. As the nature of this research project is fundamentally inter-disciplinary, the broad interpretability and versatility of the four dimensions outlined above is a necessary precondition to allow researchers, experts and academics from a wide variety of disciplines to discuss the topics from their points of view without limiting their potential interactions to colleagues from the same field. The specific items for each of the four dimensions were collected in pre-study workshops and discussions, but should be seen as a suggestion or baseline that needs to be amended or reconfigured through open questions or allowing study participants to choose their own items through an 'other' field.

## 4    Delphi Study "Digital Transformation"

The Delphi method has been widely used for exploratory studies focusing on complex, difficult to grasp topics [31]. Part of a group of foresight methods, it is based on group consensus as an evaluation measure and provides an approach to estimating future developments that, due to their complexity, defy other statistical or qualitative methods. The RAND corporation developed the Delphi survey technique in the 1950s, with the aim to provide a decision making and technological forecasting tool based on expert opinions [13]. Since then, a number of variations have been developed such as the *modified Delphi* and the *real-time Delphi* (see [17, 18]), providing evidence of the versatility and flexibility of the method. In fact, Rauch notes as early as 1979 that a large majority of studies falling under the label *Delphi study* have adapted the original, *classical Delphi* method to suit the topic of investigation, with this study being no exception to that observation. For the same reason, Linstone and Turoff even refuse to give a *"detailed and explicit definition"* of the Delphi technique in their seminal book *The Delphi Method: Techniques and Applications* [39].

Shared element to the different Delphi variations is the implementation of a structured feedback loop that allows participants to gain some notion of the groups' answers in the form of a measure of consensus for quantitative data or prepared summaries for qualitative data. For the classical approach, these data needed to be provided by the researchers during a set of *rounds*, after which all participant's answers were aggregated, summed up, and finally redistributed them before the next round of answers. Through this approach, the participants should be able to weigh the arguments brought forth by their colleagues and adapt their opinion accordingly. Facilitating a process unbiased by personality and deference to authority dictates the implementation of strict anonymity of the participants, to ensure opinions of well-known participants do not skew the results.

For this study, a variant of the Delphi method called "Real-Time Online Delphi" was implemented. Based on the real-time variant of the original Delphi as introduced by Gordon and Pease [12], this method features a round-less design in an online format, providing the participants with statistical analyses of their answer compared to the median of all other answers to any particular question after their first answer. Furthermore, this online variant allows the collection of qualitative data in the form of public comments to each question of the survey. In order to minimize the effort of conducting the study, the "SurveyJet" tool by Calibrum, Inc.[1] was used. For a detailed description and comparative evaluation of the tool and the *Real-Time Online Delphi* method, see Aengenheyster et al. [3].

### 4.1 Questionnaire and Survey Design

The questionnaire for the survey translates the framework into specific questions for the participants. Questions were collected into pages for each research question, formulated either as discrete ratings on standardized Likert scales from 1 to 5.

Each page consisted of the same question, applied to a number of items. Each question about a single item was accompanied by a comment field that instructed participants to explain their decision in more detail, allowing the collection of qualitative data and discussions between the participants.

In addition to the main content questions, demographical data about the participant's *age*, *country of residence* and self-attribution of the participant's research focus to one of the five academic disciplines *Arts*, *Humanities*, *Social Sciences*, *Natural Sciences* and *Applied Sciences* were collected.

The survey was tested in trial-runs, leading to an estimated average completion time of 30 to 45 minutes per participant and answering session.

### 4.2 Expert Selection Process

Multiple authors have stressed the importance of the selection process for the experts participating in the Delphi survey (e.g. [2, 27, 32, 36]). Ideally, participants are familiar with the topic in general and the specific areas of inquiry, while representing a diverse sample of approaches or opinions towards the topic. Thus, is it not uncommon to forgo probability sampling techniques in favour of purpose or criteria sampling, as Hasson notes:

> *"Here participants are [...] selected for a purpose, to apply their knowledge to a certain problem on the basis of criteria, which are developed from the nature of the problem under investigation."* [18, p.1010]

In the case of this study, a panel of 35 experts from a heterogeneous group of academic disciplines and fields was selected. As main criterion for participation, the participant's research focus and output had to include publications focusing on aspects, technologies or impacts of the Digital Transformation as defined in Section 2. Potential participants were identified through a literature survey of academic publications connected to or investigating aspects of the Digital Transformation, filtered by author-defined keywords, title or explicit references to the *Digital Transformation*. Special care was taken to involve experts from certain core fields with a strong connection to the topic, i.e., fields which featured multiple publications with the express focus of *Digital Transformation*, *Digitization* or *Digital Disruption*. The selected candidates were then contacted via email with a standardized invitation, explaining the process, requirements and time frame of the study, and requesting their consent to participate. Out of 118 candidates contacted, 35 responded positively, leading to an approximate response rate of 29.6%.

### 4.3 Analysis and Validation of Results

Validation of the quantitative results was sought in the form of a *measure of consensus*, specifically the quartile coefficient of dispersion *QCD*, defined as follows:

$$QCD = \frac{Q_3 - Q_1}{Q_3 + Q_1} \tag{1}$$

The QCD provides a robust measurement for comparison of the interquartile dispersion of (median) scores between each question, sections and the survey as a whole; higher QCD values are interpreted as indicative of a weaker consensus.

Additionally, the comments to each question were treated as valuable qualitative data to explain the provided answers and gain insights in the process of discussion on each of the questions. Loosely following Mayring's standard methodology for the analysis of qualitative data [25], the comments were subjected to a process of *summarization*, *inductive categorization* through keywords and *explication* where necessary. The resulting set of keywords gave insight into referenced technologies, cross-references to other questions and specific predictions or prognoses made by participants about future developments. The stability measurement for the quantitative data as described above were not used to discard answers based on a given threshold, but interpreted in conjunction with the qualitative data for any given question.

Together, the quantitative and qualitative analysis provides what Clifford Geertz called a *thick description* of the participant's answers, validity of the consensus, and culture of discussion [11].

## 5 Results

The following section discusses the results of the quantitative and qualitative parts of the study and evaluates the feedback given by the participants on the survey design and research framework. In the spirit of brevity, only a selection of items for each question are discussed.

The survey is divided into four basic questions Q1 through Q4 into the relevance or effects of the Digital Transformation in the four main dimensions defined in Figure 1.

### 5.1 Q1 – Impact on Societal Areas

The first question about the impact of the Digital Transformation on different areas of society is illustrated in Figure 2 as a divergent centred stacked bar chart, with the median participants' choice for each of the items added in as a combined axis (see [29] for a description of this visualization type for Likert-scale data). The top axis denotes the participants' median scores. The bottom axis visualizes the participant response distribution for each choice, centred around the median choice *medium impact* (with a numeric value of 3).

On average, the participants determined the impact of the Digital Transformation to be medium to strong. Especially noteworthy are the top scores for *Business / Economy* and *Industry*, reflecting the general focus of the literature as described in Section 2.

Arguments for a strong impact included one participant's opinion that *"[...] computation is largely about efficiency gains in other activities, [thus] the business impact is of higher productivity in general."* or the changes ICTs bring to production processes through automation. One participant predicts two trends in the manufacturing and service sectors:

> *"Two trends will compete in the near future: the continuation of mass–production pushing it t[o] cover almost all manufacturing and service sectors, and the re-emergence of a differentiated production focusing on different consumer profiles and different levels of quality. While the former will increase the social problems characterizing the planet, the second could support a better distribution of the creation of value, a consolidation of circular economy[, and] a more sustainable world."*

Hindering factors to a rapid transformation as seen by the participants are a diffusion of the impact due to the many different types of work which all present their own unique challenges to automation and digitalization of processes, and the observation that the majority small and medium enterprises (SMEs) were not yet sufficiently prepared for the changes brought on by the Digital Transformation.

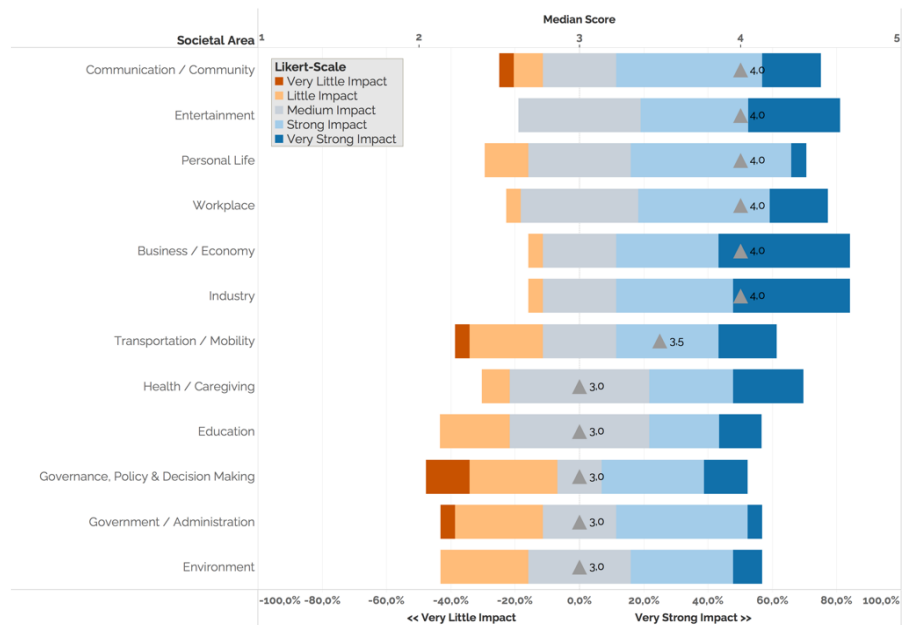

**Figure 2.** Question Q1 on the Impact on Societal Areas – Responses and Distribution

In terms of the *entertainment* industry, participants noted that new forms of distribution of media provide chances for smaller start-ups to disrupt existing business practices of large corporations, and cited the entertainment sector as a particularly technology-driven part of the economy and, consequentially, often an early adopter of new technologies such as augmented or virtual reality applications (AR/VR).

Furthermore, they predicted a continuous growth in importance of *digital games* as a medium of entertainment.

As to the future of the *workplace*, the qualitative data reveals the participants focused both on quantitative changes to employment and the way different workplaces would transform due to the Digital Transformation. As one participant sums up their arguments for a *very strong impact*:

> *"What people will do at their work-places [...] will be dramatically impacted by the digital revolution. Workplaces will change quantitatively (for the first time, even the most optimistic observers of the diffusion of digital technology admit that the number of work positions that will disappear is probably greater of those that digital technology will create) and qualitatively (whatever a person will do at her workplace, it will be always more be digitized). As a reaction to both these changes, there will be dramatic social movements at the planetary level: migrations, tensions between states, religions and ideologies, etc."*

Referencing health / caregiving, participants reiterated the strong potential for innovation due to digital technologies, but cited *"conflicting interests"* of *"large, entrenched players"* as slowing down adoption of new technologies.

Finally, the discussions about the impact of the Digital Transformation on *government / administration* and *governance, policy & decision making* revealed two separate forms of impact: on the one hand, participants predicted a slow, but continuous increase in adoption of eGovernment-applications that would affect the relationship between governments and their citizens, but cited the *"inertia"* of political systems, a *"lack of technical capacity"* and the continuing effects of the *Digital Divide* as hindering factors. On the other hand, participants saw an impact on the political landscape and democratic processes due to the wide adoption of Social Media and the related issue of Fake News.

### 5.2 Q2 – Issues affected or emerged

Question 2 asked participants to rank the importance of various issues raised by the Digital Transformation on a scale from *Not important* to *Very Important*. The quantitative results are depicted in the same form as in Q1 in Figure 3.

The participants overwhelmingly ranked the two issues *privacy* and *computer security* as *very important*. The qualitative data reveals little dissent about the dangers of the unregulated and seemingly insatiable appetite both state actors and corporations display in terms of (personal) data collection. As one participant sums up:

> *"Digital data continue to open our private lives wider and wider to the scrutiny of others. Constant surveillance is the basis of every totalitarian dream / nightmare / utopian novel. Getting this one right is crucial to our freedom, safety and happiness."*

Others take a more pessimistic stance and declare the *"fight [...] lost already"*, call the amount of data collection as it happens now *"scary"* or note the lack of good models of *"[...] how online privacy should work"*, although they mention the *"Right to be forgotten"* and the European Union proposal for General Data Protection Regulation (GDPR, see [8]) as steps in the right direction.

Similarly, participants note both the importance of computer security and their belief that security breaches and hacking will increase, with one participant going as far as declaring that *"there is no such thing"* as computer security and another simply stating *"no computers = security[,] computers = no security"*.

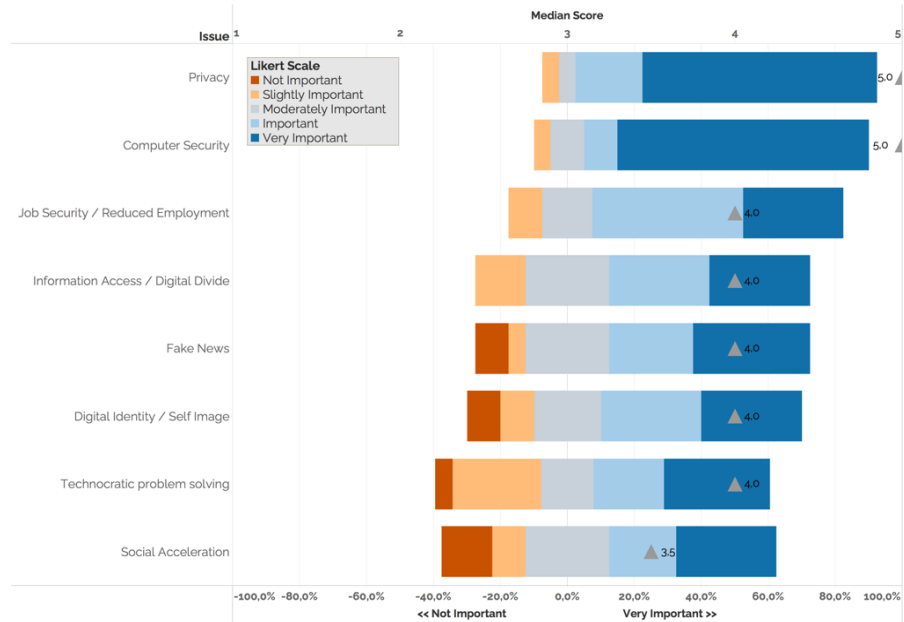

**Figure 3.** Question Q2 on Issues Affected or Emerged – Responses and Distribution

Discussing the issues of *job security / reduced employment*, opinions were diverging. While some saw this issue as *very important*, citing the profound impacts reduced employment can have on societies and automation as a driving factor, others doubted the projected job losses or whether any reasonable prediction could be made in this question.

Finally, on the question of *Fake News*, participants acknowledge the existence and real-world impact of the phenomenon as a destabilizing factor, with one participant declaring Fake News *"Impossible to stop; devastating in impact."*. Others argue that the spreading of false or misleading information as a means of manipulation is an issue that predates modern technology by centuries, but concede that the advent of digital communication technologies has amplified the effect. One participant ventures to situate the issue within a greater societal scope, quoting *"relativism, disbelief in science [and] filter bubbles"* as the larger phenomena encompassing *Fake News*.

### 5.3 Q3 – Impact on Global Challenges

Looking at the potential impact the Digital Transformation will have on the list of *global challenges*, the median scores on a scale from *strong negative effect* to *strong*

*positive effect* are, at first sight, much less conclusive than the results of questions Q1 and Q2. For the overview of the quantitative results refer to Figure 4.

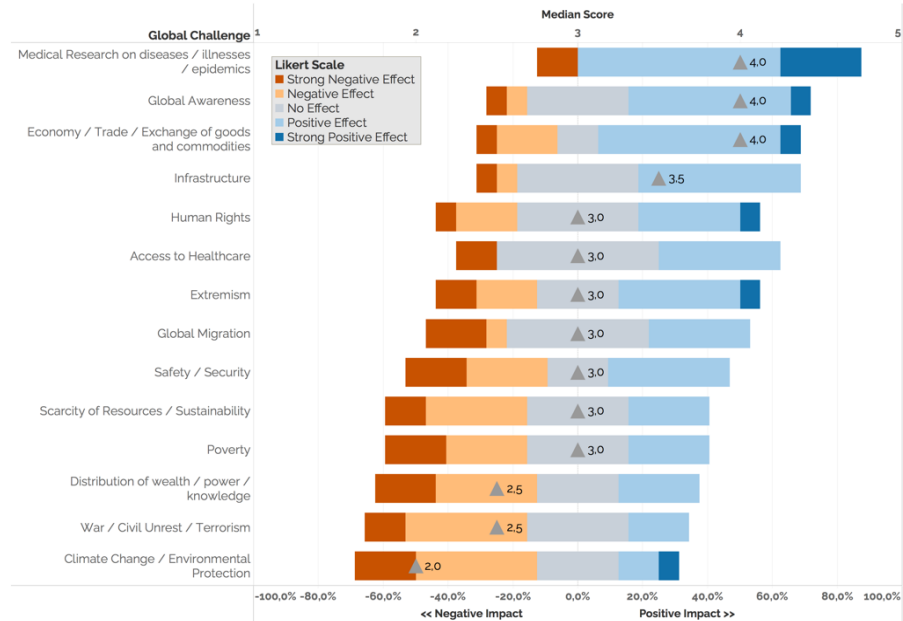

**Figure 4.** Question Q3 on the Impact on Global Challenges – Responses and Distribution

The only *positive* median scores can be observed for the challenges *medical research, global awareness* and *economy / trade*. The qualitative data reveals the participant's positive outlook specifically for *medical research*, citing advances in medical health tracking and monitoring in general (at the expense of privacy), *"digital tools"* and *"increased data processing capability"* as the drivers. Nevertheless, one participant stresses the political will as a key factor for success in this global challenge, while another points out the limitations of digital tools by pointedly stating that *"[... medical] research is still primarily dependent on in vivo or in vitro, not in silico."*

Beyond that, the average ranking for the other challenges remains largely within the median value of no effect; given the distribution of answers as depicted in Figure 4, this is less due to an overwhelming consensus that the Digital Transformation will not have an impact than more due to the diverging opinions on whether there will be positive or negative effects. For multiple challenges, including *infrastructure*, *extremism* and *human rights*, the participants stated different potential positive or negative effects that, overall, might cancel each other out or be too complex to predict. One participant's comment on the challenge of extremism illustrates this stance particularly well:

*"Digital techniques will be used on both sides, and will mostly cancel out."*

Nonetheless, the fact that the qualitative data contained examples or scenarios for both positive and negative effects of the Digital Transformation for all listed challenges suggests that, while difficult to pinpoint, that the Digital Transformation carries the potential for a strong overall impact in either direction of the scale.

### 5.4 Q4 – Technologies Involved

Question Q4 focuses on the overall impact different *digital technologies* will have in the next five years, providing a snapshot view of current trends as seen by the participants. The median scores and distribution of choices on a scale from *very little impact* to *very strong impact* are depicted in Figure 5.

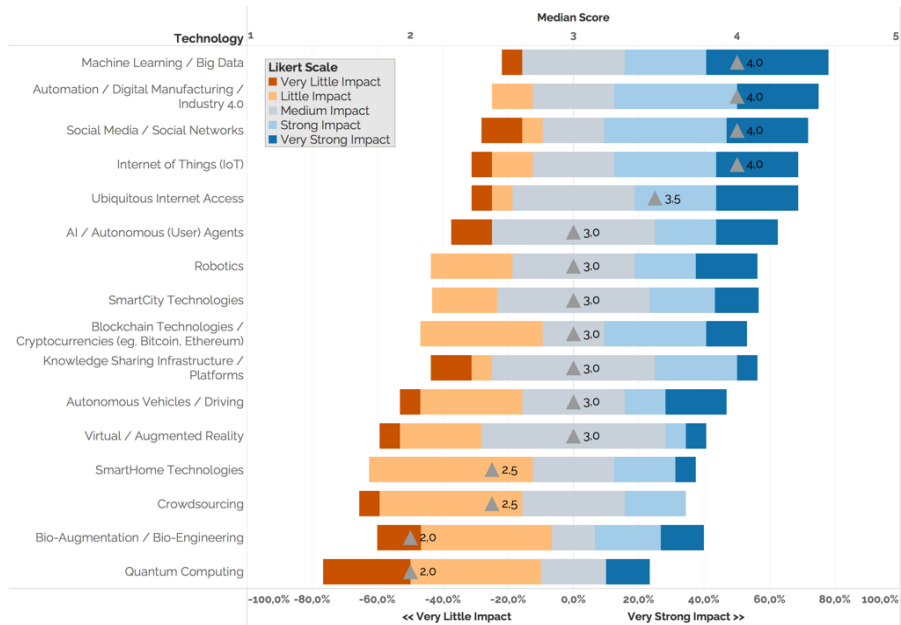

**Figure 5.** Question Q4 on Technologies Involved – Responses and Distribution

The median scores submitted by the participants show some clear trends, but the supplemental qualitative data also shows some controversial discussion points for specific technologies.

The first two most impactful technology complexes as selected by the participants are *machine learning / big data* and *automation / digital manufacturing / industry 4.0* with a median score of *strong impact*. For *machine learning*, participants stated an already existing strong impact, and saw that still increasing in the coming years, but also related the topic to the aforementioned issue of *privacy* as carrying strong social implications. As with many of the technologies in this questions, participants were eager to point out that, beyond the factual capabilities, *"organizational and political factors"* would be determining the impact of *machine learning* in the years to come.

Likewise, the participants explained their choices for a *strong impact* of *automation* in accordance with their previously stated implications of automation technologies for *employment* and, subsequently, *poverty*. A dissenting voice referred to the three terms as *"buzzwords"*, explaining their choice for a small impact.

One of the most controversial discussion occurred about the future impact of *blockchain technologies / cryptocurrencies*. On the one hand, participants argued for a separation of the two, foreseeing different impact for each of them:

*"I think blockchain might turn out to be important for its disintermediating effects. Not convinced that cryptocurrencies are much more than a device for illegitimate exchange."*

Others took a clear position in favour of a very strong impact, reiterating the importance of blockchain technologies *"[...] [a]cross industries and not just for cryptocurrencies"*. A third opinion ascribing *little impact* was given by a participant mentioning social engineering as a potential vector of attack against the security measures inherent in blockchain technologies, who also pointedly drew a comparison between the high price of Bitcoin at the time and the Dutch *tulip mania* in the mid-17$^{th}$ century by stating *"Tulips were not in fact a stable store of value or medium of exchange"*.

## 6     Conclusions

The Digital Transformation is ongoing and far-reaching – not only in business and economic processes, but also in most other areas of society on several levels. Changes in societal structures, values and perceptions can increasingly be observed. It is imperative we first understand what Digital Transformation means by considering different perspectives to the concept and its impacts. Beyond understanding, the ability to create robust projections of future developments is crucial to political and business leaders in order to adapt and focus their efforts for research, policy and resource allocation.

In this paper, we tried to approach this problem with an Online Real-Time Delphi study. By identifying the context of the Digital Transformation and conceptualizing its dimensions we created a common rationale and language for experts to discuss emerging trends and projections. The results showed the potential of a large number of subjects related to Digitalization and Digital Transformation in the next five years. The biggest attention should be given – as the experts of the study agreed upon – to the societal areas *economy*, *industries* and *health care* as well as to problematic issues such as *privacy* and *security*. Technologies like *machine learning*, *automation*, *social media* and *IoT* are still the most likely to exert a strong impact in the years to come. The goal must be to explore the positive potential of technologies and emerging technological approaches. This can be achieved by establishing research practices that facilitate a critical reflection to the development and introduction of technologies into societal areas. It should be possible to capture and analyse emerging problems caused by the use of new technologies and, beyond that, prevent the increase of problems and issues for society at large.

Furthermore, this paper also provides evidence that studies following the Delphi method are valuable tools for qualitative capturing of emergent developments in our society. They help frame the context of investigations, document experts' opinion and arguments in the given specific context, and assure an exchange among experts towards a convergence of different directions. It is clear to us that we have to continue gathering experts' opinions on new developments in the future and – based on the well-analysed input of experts – with discussions among them and other stakeholders presumably being exposed to the new developments.

Besides demonstrating the suitability of the Delphi methodology, this study also encourages continued research into these topics through a broad, socio-technical approach that is decidedly humans-first. Through a multi-faceted, exploratory data collection process we gathered input from discussions, visits to different organisations focused on the Digital Transformation and a survey of existing literature from many different fields. This allowed us to construct a robust research framework that will be an important tool for studying the Digital Transformation not simply as a technological phenomenon but rather as a more complex trajectory of technologies, (global) issues and societal areas. As the rapid transformation of society in an increasingly digitized world continues, a balanced investigation into technologies and innovation, the economy and society at large remains a major challenge that requires a concerted effort of the Technical Sciences, Social Sciences and Humanities. Our future work will focus on engaging this topic through such an interdisciplinary approach to further our understanding of Digitalization and the Digital Transformation of society.